# Fat Tails Quantified and Resolved:
# A New Distribution to Reveal and Characterize
# the Risk and Opportunity Inherent in Leptokurtic Data

Lawrence R. Thorne
http://lawrencerthorne.com

**Abstract**

I report a new statistical distribution formulated to confront the infamous, long-standing, computational/modeling challenge presented by highly skewed and/or leptokurtic ("fat- or heavy-tailed") data. The distribution is straightforward, flexible and effective. Even when working with far fewer data points than are routinely required, it models non-Gaussian data samples, from peak center through far tails, within the context of a single probability density function (PDF) that is valid over an extremely broad range of dispersions and probability densities. The distribution is a precision tool to characterize the great risk and the great opportunity inherent in fat-tailed data.

**Keywords**: fat tails, heavy tails, leptokurtosis, skew, power-law, risk, value at risk, kernel density; self-similarity

## 1. Introduction

In the annals of statistical analysis, there has been a longstanding reliance on the applicability and power of the normal, or Gaussian, distribution. Its very name implied that it was the norm, the standard by which most data should be evaluated. This implication was reinforced in financial circles when the Gaussian was enshrined in Markowitz' modern portfolio theory, then in the Nobel-Prize-winning Black & Scholes Model; thereafter, new analytical protocols were routinely based on "the Gaussian assumption." But as the demands of quantitative analysis became increasingly sophisticated, instances in which the Gaussian distribution proved inadequate proliferated, and whispered mentions of worrisome "fat tails" began to creep into water-cooler discussions and the professional literature. In recent years, the whispers have swollen to an outcry. An entire vocabulary now exists to describe the unusual or "extreme events" that occur in the tails of leptokurtic probability distributions and that cannot be modeled adequately by the bell-shaped Gaussian distribution. Worried pundits mutter darkly about "outliers" or poetically about "black swans." And news stories reminiscent of those surrounding the catastrophic 1998 collapse of the investment firm Long-Term Capital Management continue to accumulate, documenting the dangers of underestimating either the likelihood or the severity of "fat tail events."

In response to the growing awareness of the severe limits of Gaussian validity, serious scholars have focused their attention on the search for alternatives. Innovative statistical distributions have sprung up almost as quickly as extreme-event slang; and some of these distributions have been significant developments that met particular analytical objectives. But no one of the new distributions has proven to be a generalized distribution suitable for use everywhere and anytime the Gaussian is inadequate. So frustrated data crunchers have been reduced, at times, to classifying extreme data points as outliers and simply ignoring them; desperate quants have resorted to generating esoteric approximations. Creativity has abounded. And fat-tail risk has continued to hide in plain sight.

The foundation of the distribution announced here is a well-known and very basic construct. Its implementation, however, is fresh, forceful and potentially important.

**2. Practical characteristics of the new Thorne Distribution**

- The distribution models non-Gaussian data samples, from Gaussian center to power-law-like tails, within the context of a single PDF.

- The distribution provides a complete and accurate description of tail behavior, including the higher moments of skew and kurtosis.

- The distribution extracts, from time series or other data, highly accurate probability density functions (PDFs) that are valid over a broad range of dispersions and densities. For the unconditional PDF of the log-price return of S&P 500 tick data, for instance, the Gaussian Law is valid for a total range of only about five (5) standard deviations centered around the mean; the new distribution covers 85 standard deviations. The Gaussian is limited to tail densities greater than $10^{-2}$; but the new distribution extends to $10^{-7}$. (Please see Figures 1 and 2 below.)

- The distribution requires many times fewer data points to produce a high-quality PDF than do conventional methods such as the classical histogram approach or the more advanced Sheather-Jones kernel-density method.

- Even when applied to small, leptokurtic data sets, the distribution yields tail characterizations that are statistically superior to those produced by Extreme Value Theory (EVT), which is the input most commonly used for Value-at-Risk (VaR) and Expected Shortfall methodologies.

- The distribution is highly flexible. It adapts to data from a wide range of sources and can accurately describe unimodal, bimodal and multimodal PDFs.

- The distribution can be computed rapidly. This, together with the fact that it can function with very small data sets, permits rapid updating of time-series PDFs.

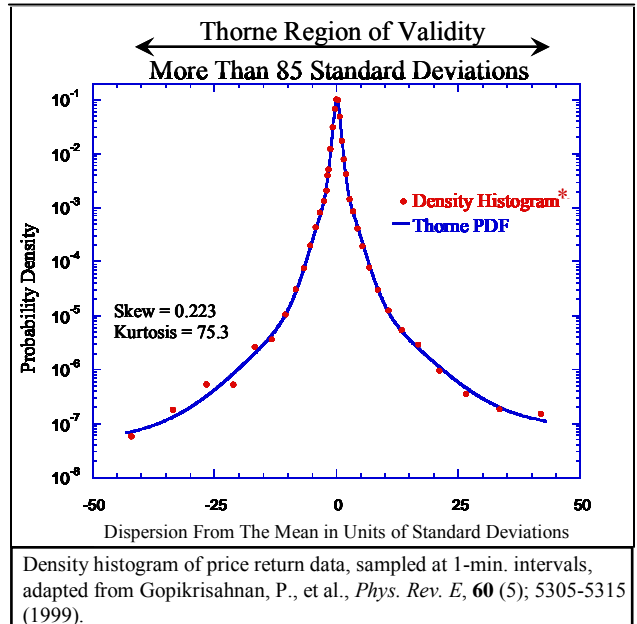

Figure 1. Thorne Distribution characterizes entire domain of S&P 500 density histogram accurately.

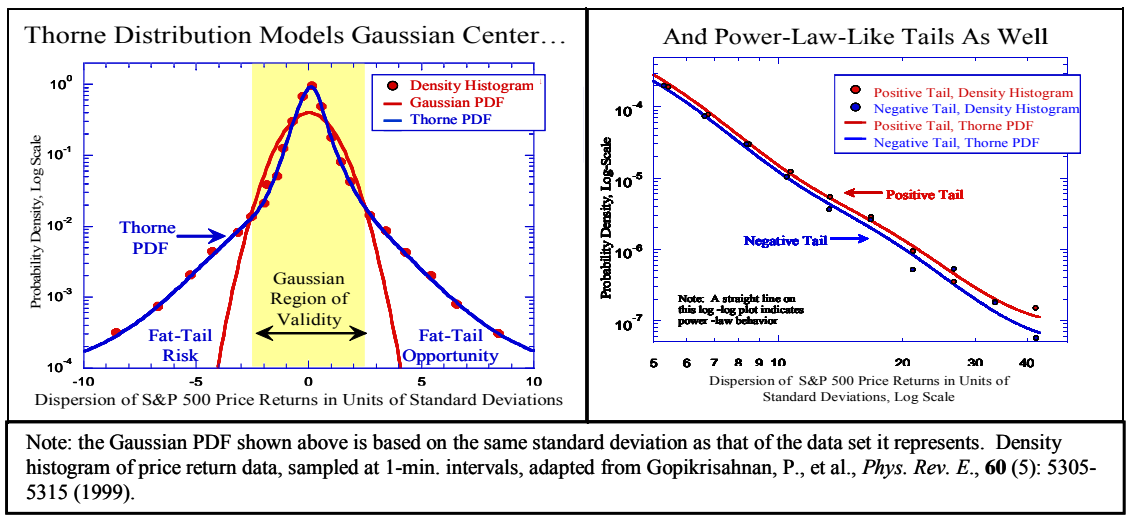

Figure 2. Expanded views of new PDF for the center and tail regions of the specified data set illustrate excellent fit.

## 3. Need for a new distribution

To deal correctly with the varied, changing curvatures of leptokurtic PDFs, one would—until now—have to cobble together a patchwork of several distributions, each describing a limited section of the overall distribution. A Gaussian distribution, for example, could be used to characterize the center of a leptokurtic distribution; power-law distributions would work for intermediate tail regions; and generalized extreme-value distributions would describe the far tails. This approach mixes distributions from radically different families, however, and it is hugely problematic. Within it, there is no objective basis for defining the specific section of the PDF that any, one distribution will dominate; "splice" points between distribution regions commonly introduce discontinuities in the PDF; and parametric values become dubious. A data set thus analyzed will have a segmented, rather than a continuous, distribution. Also, since they do not work in concert, patched-together distributions do not yield a self-consistent, complete understanding of the fundamental nature of the PDF of a data set. In contrast, the Thorne Distribution characterizes a data set from peak center through far tails within the context of a single PDF.

Perhaps even more importantly, the new distribution is flexible enough to handle the full spectrum of fat-tail shapes. There are a number of distributions, like the Cauchy, that do represent a fat-tail shape; but each of them produces only one, single, distinctive tail shape characterized by slowly varying, monotonically-decreasing curvature progressing toward the far tails. In other words, each of these "fat-tailed distributions" is suitable only for data distributed in a very specific pattern. The new distribution, however, is flexible enough to model an extremely wide range of tail shapes, including even platykurtic ones.

| CHARACTERISTICS | DISTRIBUTION (PDF) NAMES | | | | | | |
|---|---|---|---|---|---|---|---|
| | Cauchy | Pareto | Student-t | Levy | Stretched Exponential | Generalized Extreme Value | Thorne |
| Can model peak center through far tails with consistent accuracy over > 50 standard deviations | | | | | | | ✓ |
| Can model multiple leptokurtic tail shapes | | | | | | | ✓ |
| Can revert to Gaussian shape in extreme tails | | | | | | | ✓ |
| Can model platykurtic tail shapes | | | | | | | ✓ |
| Can model power-law-like fat tails | ✓ | ✓ | | ✓ | ✓ | ✓ | ✓ |
| Can model a Gaussian or quadratic center | ✓ | | ✓ | ✓ | ✓ | | ✓ |
| Can model extreme (> 15) leptokurtosis | ✓ | ✓ | ✓ | | ✓ | ✓ | ✓ |
| Can model higher (> 2) moments | | ✓ | ✓ | | ✓ | ✓ | ✓ |
| Can model symmetric distributions | ✓ | | ✓ | | ✓ | | ✓ |
| Can model skewed distributions | | ✓ | | ✓ | | ✓ | ✓ |

Table 1. The characteristics of the Thorne Distribution are comprehensive.

## 4. Description of the new distribution

The basic Thorne Distribution is a log-log tiered Guassian—the log-sum of Gaussians with log-transformed, positive variates (or differences of such variates). It is constructed to describe the PDF of leptokurtic data; it also handles platykurtic data. It is the normalized, finite, exponentiated, weighted sum of multiple Gaussian distributions (hereafter, "component Gaussians") that fits the log-transformed PDF of the log-transformed variates of a data set. It meets the formal definition of a distribution in that it integrates to one (1) and has no negative values. Actually, it constitutes a multi-scale distribution of distributions. The new distribution is parametric, continuous, non-elliptical and self-similar. It has relatively low statistical roughness; it is not stable. And it contains, as a sub-set, the familiar Gaussian distribution itself in the log domain.

4.1. Weighted sum of Gaussian distributions

Summing Gaussians is not a new idea; it is used in mixture models, for instance. Using the sum of multiple Gaussian distributions either to characterize the probability density function of fat-tailed data or to characterize the log transform of such basic data is not unheard of. The latter is equivalent to a sum of log-normal distributions for untransformed variates. But the Thorne Log-Log Tiered Gaussian Distribution is unique in that it uses the normalized sum of multiple Gaussians to characterize the logarithm of the PDF of log-transformed variates. This PDF is not equivalent to using the sum of log-transformed Gaussians because the log of the sum of functions is not equivalent to the sum of the log of the functions; this is because the log transformation is not a distributive operation for summation, as is well known. And this is an all-critical difference! (See Eq. 2.)

4.2. Dual use of log transformations

The adaptability of the new distribution stems chiefly from its dual use of log transformation to reduce the dynamic range of, first, the raw variates of the subject data set itself, and, second, the range of its underlying PDF. For time series, the first log transformation of the variate of the data set also transforms the raw data into a type of normalized space that is described by additive stochastic processes rather than by multiplicative processes. And this log transformation, which creates the normalized space, converts the raw variates into a self-referential metric or numeraire. The second use of the log transformation—taking the logarithm of the PDF—reduces its relative vertical scale and makes the PDF amenable to characterization by a weighted sum (weights > 0) of component Gaussians of varying means and widths.

Both of these logarithmic transformations are essential. The first, taking the natural log transformation of the always-positive variates of a data set, is routine in statistical analysis. But the second, taking the log transformation of the histogram density that is determined from the basic data set, is highly unusual. And it is this second transformation that provides inherent logarithmic weighting for the fitting process later used to extract the parameters (weight, width and mean) for each of the component

Gaussian distributions that comprise the larger Thorne PDF. Otherwise, the default, uniform weighting of the fitting process would give undue emphasis to the center of the distribution and virtually ignore the very small, but very important, probability densities in both far tails. The second transformation is also advantageous because it effectively compresses the probability density range of the histogram and thereby gives heightened statistical weight to points in the tail of the new distribution. This compression of the range facilitates expanding the usable support, or domain, of the Thorne PDF beyond typical limits.

4.3. Formulation of the distribution

The Thorne probability density function is given by:

$$f(x) = \exp\left[\sum_{i=1}^{n} \frac{w_i}{\sigma_i \sqrt{2\pi}} \exp\left[-\frac{(x-\mu_i)^2}{2\sigma_i^2}\right]\right] - 1 \qquad (1)$$

or, in the log domain, as a sum of weighted Gaussians:

$$\log_e[f(x)+1] = \sum_{i=1}^{n} \frac{w_i}{\sigma_i \sqrt{2\pi}} \exp\left[-\frac{(x-\mu_i)^2}{2\sigma_i^2}\right] \qquad (2)$$

Where:
$f(x)$ = Thorne Distribution or PDF
$x$ = independent variable, i.e., typically the natural logarithm of variate values or the differences between them, as in log return data
$i$ = index variable for the $i^{th}$ Gaussian
$n$ = total number of component Gaussians
$w_i$ = weight factor of the $i^{th}$ Gaussian ($w_i > 0$; $w_1 < w_2 < \ldots w_n$)
$\mu_i$ = mean, or location, of the $i^{th}$ Gaussian
$\sigma_i$ = standard deviation, or scale, of the $i^{th}$ Gaussian ($\sigma_1 < \sigma_2 < \ldots \sigma_n$)

4.4. Normalization of the new distribution

The basic probability density function, as given in Eq. 1, is normalized automatically when its parameters are determined by a fit to a normalized density histogram. Since such normalized histograms are so commonly used, Eq. 1 does not specify an explicit normalization term. Such a term, however, can be calculated easily by evaluating the integral, in the equation below, numerically. Notably, the minus one (1) term in this equation is critical to the successful convergence of the integral. Values even slightly different from one (1) do not permit convergence over the domain of support ($-\infty, \infty$).

$$N = \int_{-\infty}^{\infty} \left[ \exp\left[ \sum_{i=1}^{n} \frac{w_i}{\sigma_i \sqrt{2\pi}} \exp\left[ -\frac{(x-\mu_i)^2}{2\sigma_i^2} \right] \right] - 1 \right] dx \qquad (3)$$

Where N is the divisor that normalizes the PDF.

Note: Eq. 3 can also be used to normalize the Thorne PDF of an optimized, but not normalized, density histogram.

4.5. Normalization of a special case within the new distribution

In rare, limited cases, a user might wish to use a single component Gaussian PDF based on the new distribution. This is not a complete Thorne Distribution, which must consist of two (2) or more component Gaussians; nonetheless, it is a new distribution and can be a useful tool. For clarity's sake, let's call this PDF a Truncated Thorne Distribution. Normalizing it requires the use of a new mathematical constant that may be unique to this distribution; it is not included in the literature among well-known mathematical constants. This Thorne Constant ensures that the subject PDF integrates to one (1). The PDF is symmetric and mildly leptokurtic. It is specified by the normalized equation below. Note that the argument of the inner exponential function therein is a normalized Gaussian distribution multiplied by $\sigma$ and that $1/\sigma$ is required to normalize the PDF.

$$f_{truncated}(x) = \frac{1}{C_T \cdot \sigma} \left[ \exp\left[ \frac{1}{\sqrt{2\pi}} \exp\left[ -\frac{(x-\mu)^2}{2\sigma^2} \right] \right] - 1 \right] \qquad (4)$$

Where:
$f_{truncated}(x)$ = truncated Thorne PDF
$C_T$ = Thorne Constant, 3.697252480597963…, and is given by the equation:

$$C_T = \int_{-\infty}^{\infty} \left[ \exp\left[ \frac{1}{\sqrt{2\pi}} \exp\left[ -\frac{(x-\mu)^2}{2} \right] \right] - 1 \right] dx \qquad (5)$$

Note: this integral can be evaluated numerically for normalization purposes.

4.6. Role of the component Gaussians

Specifically, the Thorne Distribution is the exponentiated sum of two (2) or more— usually three (3) or more, with leptokurtic data—component Gaussians. (The fit illustrated in Figure 1 above, which covers 85 standard deviations, was achieved with only three (3) component Gaussians.) Overall, the total number of component Gaussians employed is determined by the detailed characteristics of an individual data set and by the objectives of the PDF's user.  The distribution is extremely flexible and adapts easily to varying needs.  A PDF with large leptokurtosis ($\kappa > 10$) and/or skew ($|s| > 1.5$) or with a demand for high accuracy over a wide span of standard deviations from the mean (i.e., with wide dispersion and support) will necessitate using more component Gaussians than will a thin-tailed, symmetric PDF described across only a modest number of standard deviations. The mean values, i.e., the location parameters, of the component Gaussians are routinely similar but not equal in unimodal data sets; in multimodal data sets, the means may vary widely.  Kurtosis determines the width of the widest component Gaussian.  Mean values shift positively or negatively depending on the positive or negative skew in a data set. Remarkably, however, the resultant PDF of exponentiated, summed Gaussians remains smooth and continuous for a wide range of means, weights and widths of component Gaussians.

Separate, single-component Gaussians dominate in the center and tail regions of the subject PDF.  The center—within two (2) standard deviations of the mode/peak—is determined primarily by the component Gaussian with the narrowest width; the tails are determined primarily by the component Gaussian with the broadest width.  The intermediate, convex regions of the PDF are characterized by all but the most narrow of the component Gaussians.  Surprisingly, these intermediate regions in the Thorne Distribution often show power-law-like behavior, which suggests scale invariance in the regions.  Yet happily, unlike true, power-law dependence, $f(x) = x^{-(1+\alpha)}$, that produces inappropriately high density in extreme tails, the Thorne Distribution ultimately reverts to Gaussian behavior in the most extreme tails.  Distributions such as the stable Levy or the unstable, truncated Levy do not revert, and neither do the Cauchy, Pareto or stretched-exponential distributions.  It is an often overlooked reality that power laws are strictly appropriate in only the intermediate-tail regions of the PDFs of most real-world data sets. In such PDFs, the densities in extreme-tail regions decrease faster than any power law [1].  In sum, the Thorne Distribution exhibits robust flexibility that allows it to describe probability density accurately across the full support of an entire data set, regardless of how leptokurtic the data set may be.

4.7. Numbers of parameters of the distribution

The number of component Gaussians employed in the new distribution determines the number of free parameters that must be considered, obviously.  Conventional wisdom favors distributions that use as few parameters as possible, i.e., parsimonious formulations.  Superficially, it would appear that the multiple, component Gaussians of the Thorne Distribution would introduce many parameters, since there are three (3) parameters for each component Gaussian—weight, width and mean.  This is what the basic equation for the new distribution shows.  But it is often possible to reduce this number when there is a functional relationship between the weights and the widths of the component Gaussians.  The component-Gaussian parameters of the Thorne Distribution may be interdependent; surprisingly, their weights and widths may be linearly related; and the vector lengths between successive Gaussians may be fixed.  A linear relationship between widths and weights, together with fixed-ratio vector lengths, can be used to create a recursive relation, as illustrated in the following section.

4.8. Reducing the number of parameters in the distribution

In situations in which the linear weight-width and the vector-length ratio relationships hold, the new distribution depends on only four (4) parameters, regardless of the number of component Gaussians. (A plot of component-Gaussian weights vs. widths will reveal whether or not the linear relationship exists.) Those four (4) parameters can be either the weights and widths of the first two (2) component Gaussians or a set of constants that can be derived from the specified weights and widths. One such choice of constants would be: the width of one (1) Gaussian, the slope and intercept of the linear relationship of widths to weights, and the constant describing the interval ratio of vector lengths between successive Gaussians. Using the chosen constants permits one to generate the third and all successive Gaussians, which reduces the number of requisite parameters for the total distribution significantly. Six (6) component Gaussians in a symmetric PDF, (i.e., where all component Gaussians have the same mean value) for instance, would ordinarily require twelve (12) parameters. Using the specified constants would reduce this total to four (4). If the PDF was not symmetric, an additional parameter—the mean—would be added for each component Gaussian.

The unexpected discovery of a linear relationship between the weights and widths of the component Gaussians of many data sets (such as those of log-price-return PDFs) is a rather dramatic one. An analysis of selected S&P 500 data sets illustrates it well. Each point on Figure 3 below represents the weight and width parameters for one (1) of the three (3) Gaussians employed; and all fall precisely on a straight line. One would not have expected this simplifying result *a priori*.

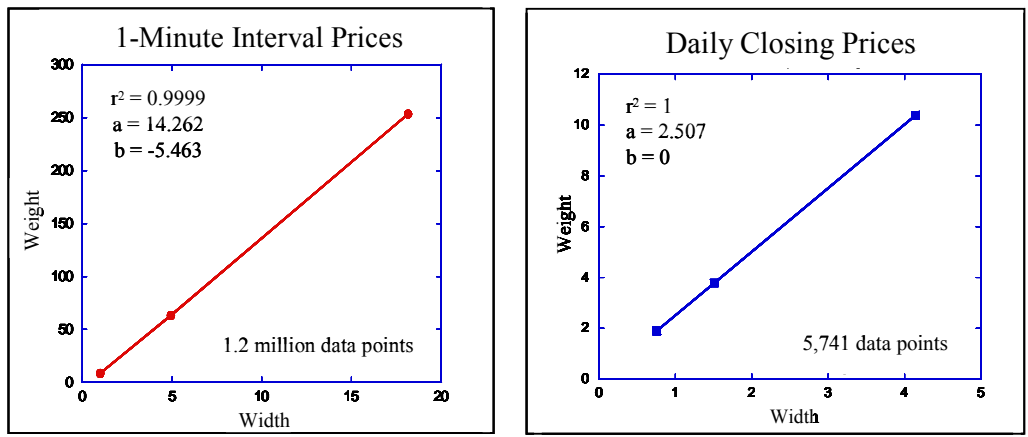

Figure 3. Component-Gaussian weight-width relationship for S&P 500 data shows unexpectedly high linearity.

Note that the three (3) points on each plot divide each line into two (2) line segments. The point coordinates for the first plot are {(0.98951, 8.6495), (4.9413, 63.184), (18.165, 253.60)}; for the second plot, they are {(0.75463, 1.8916), (1.5102, 3.7854), (4.1436,

10.386)}. Significantly, the ratio of the longer segment length to the shorter segment length on both lines is remarkably similar—3.488:1. The segment ratio of the one-minute-interval data is only 0.16% larger than that for the daily data. This ratio remains constant over a wide range of time scales and is independent of the component-Gaussian mean values and independent of skew.

The equation for the linear weight-width relationship shown is:

$$w_i = a \cdot \sigma_i + b \tag{6}$$

Where:
$w_i$ = weight of the *i-th* Gaussian
$a$ = slope (constant)
$\sigma_i$ = width (standard deviation) of the *i-th* Gaussian
$b$ = intercept (constant)

4.9. Stability of the distribution

The Thorne Distribution does not satisfy the classical definition of stability because it does not meet the requirements of the sum law of independent samples. And current quantitative thought, in the spirit of the theory of Occam's Razor, favors the use of stable distributions because they are well characterized, parametrically parsimonious and mathematically convenient. According to this thinking, any particular system can be fully represented by a single, well-chosen, stable distribution. But we now know this to be magical thinking in too many instances. Furthermore, along the way to the new distribution, I concluded that many dynamically-changing stochastic systems (such as a time series of incremental changes) cannot, in fact, be represented by a stable distribution for anything but a brief moment in time. Over a broad time span, a stochastic system may roll through a number of stable states [2] and can, therefore, be described accurately only by a distribution also capable of dynamic change, or, perhaps, by a distribution of distributions—such as the log of summed Gaussians of the Thorne Distribution.

Also, real (as opposed to synthetic) data are frequently characterized poorly by stable distributions—or even by distributions stable in their asymptotic limits, i.e., distributions that converge slowly to a stable distribution—because there are often insufficient data points in such a distribution for it to reach its asymptotic limit fully before the intrinsic stochastic process from which the data were derived changes to produce a different, perhaps closely related distribution or before a regime change occurs. In such cases, using an unstable distribution is preferable to using the stable, asymptotic-limit distribution of an unconverged data set.

**5. Statistical characteristics of the Thorne Log-Log Tiered Gaussian Distribution**

The statistical characteristics of the Thorne Distribution are given explicitly by the following equations, which are formulated as an exponentiated, weighted sum of single Gaussians.

PDF: $$\frac{1}{N}\left[\exp\left[\sum_{i=1}^{n}\frac{w_i}{\sigma_i\sqrt{2\pi}}\cdot\exp\left[-\frac{1}{2}\cdot\frac{(x-\mu_i)^2}{\sigma_i^2}\right]\right]-1\right]$$

CDF: $$\frac{1}{N}\int_{-\infty}^{x}\left[\exp\left[\sum_{i=1}^{n}\frac{w_i}{\sigma_i\sqrt{2\pi}}\cdot\exp\left[-\frac{1}{2}\cdot\frac{(y-\mu_i)^2}{\sigma_i^2}\right]\right]-1\right]dy$$

Support: $[-\infty, \infty]$

Mean: $$\left[\frac{1}{N}\int_{-\infty}^{\infty}x\cdot\left[\exp\left[\sum_{i=1}^{n}\frac{w_i}{\sigma_i\sqrt{2\pi}}\cdot\exp\left[-\frac{1}{2}\cdot\frac{(x-\mu_i)^2}{\sigma_i^2}\right]\right]-1\right]dx\right]$$

Standard Deviation: $$\left[\frac{1}{N}\int_{-\infty}^{\infty}(x-\mu_i)^2\cdot\left[\exp\left[\sum_{i=1}^{n}\frac{w_i}{\sigma_i\sqrt{2\pi}}\cdot\exp\left[-\frac{1}{2}\cdot\frac{(x-\mu_i)^2}{\sigma_i^2}\right]\right]-1\right]dx\right]^{1/2}$$

Skew: $$\frac{N^{\frac{1}{2}}\cdot\int_{-\infty}^{\infty}(x-\mu_i)^3\cdot\left[\exp\left[\sum_{i=1}^{n}\frac{w_i}{\sigma_i\sqrt{2\pi}}\cdot\exp\left[-\frac{1}{2}\cdot\frac{(x-\mu_i)^2}{\sigma_i^2}\right]\right]-1\right]dx}{\left[\int_{-\infty}^{\infty}(x-\mu_i)^2\cdot\left[\exp\left[\sum_{i=1}^{n}\frac{w_i}{\sigma_i\sqrt{2\pi}}\cdot\exp\left[-\frac{1}{2}\cdot\frac{(x-\mu_i)^2}{\sigma_i^2}\right]\right]-1\right]dx\right]^{3/2}}$$

Kurtosis: $$\frac{N\cdot\int_{-\infty}^{\infty}(x-\mu_i)^4\cdot\left[\exp\left[\sum_{i=1}^{n}\frac{w_i}{\sigma_i\sqrt{2\pi}}\cdot\exp\left[-\frac{1}{2}\cdot\frac{(x-\mu_i)^2}{\sigma_i^2}\right]\right]-1\right]dx}{\left[\int_{-\infty}^{\infty}(x-\mu_i)^2\cdot\left[\exp\left[\sum_{i=1}^{n}\frac{w_i}{\sigma_i\sqrt{2\pi}}\cdot\exp\left[-\frac{1}{2}\cdot\frac{(x-\mu_i)^2}{\sigma_i^2}\right]\right]-1\right]dx\right]^{2}}$$

Parameters: $$N = \int_{-\infty}^{\infty}\left[\exp\left[\sum_{i=1}^{n}\frac{w_i}{\sigma_i\sqrt{2\pi}}\cdot\exp\left[-\frac{1}{2}\cdot\frac{(x-\mu_i)^2}{\sigma_i^2}\right]\right]-1\right]dx$$

Where:

$w_i$ = weighting fraction for the $i^{th}$ component Gaussian
$x$ = variate value
$y$ = dummy variable of integration
$\mu_i$ = mean of the $i^{th}$ component Gaussian
$\sigma_i$ = standard deviation for the $i^{th}$ component Gaussian (square root of the second central moment)

| Tail Properties: | The distribution produces a wide range of tail curvatures, including power-law-like shapes with tail exponents that can be either integer or fractional; in the very extreme tails, the distribution always reverts to Gaussian. |
|---|---|
| Statistical Roughness: | The statistical roughness of the new distribution is dominated by the roughness of its narrowest component Gaussian. The roughness, as judged by its integrated, squared, second derivative, is much closer to the roughness of a simple Gaussian (very smooth) than to that of a log-normal distribution (very rough); it is typically one (1) to five (5) times that of a Gaussian. The comparatively low roughness of the distribution facilitates the extraction of tail densities and the determination of data-set PDFs in general. |

## 6. Thorne Stochastic Differential Equation and its solution

The Thorne Stochastic Differential Equation and its solution are the fundamental link between stochastic processes and the basic Thorne Distribution. The equation is formulated as a weighted sum of the Ito stochastic differential equations that represent the component Gaussians of the Thorne Distribution. Each individual Ito process includes a drift coefficient, $\mu$, and a diffusion coefficient, $\sigma$; but they are driven by a common Wiener process, $W$. The values of $dX$ follow the Thorne Distribution.

The non-normalized Thorne Stochastic Differential Equation for the log domain is:

$$dX = \sum_{i=1}^{n} w_i [\mu_i \, X \, dt + \sigma_i \, X \, dW] \qquad (9)$$

Where:
$dX$ = incremental change in variate $X$
$w_i$ = weight factor of the $i^{th}$ Gaussian
$\mu_i$ = drift coefficient of the the $i^{th}$ Gaussian
$X$ = stochastic variate
$dt$ = time derivative
$\sigma_i$ = diffusion coefficient of the the $i^{th}$ Gaussian
$dW$ = $\sqrt{dt} \, N(0,1)$, where $N(0,1)$ is a Gaussian random number

The solution of this equation is straightforward, as shown below. It is this solution equation that generates stochastic paths for simulating continuous-time stochastic processes whose applications might be as varied as simulating asset-price histories in finance or as super-diffusion phenomena in physics.

The solution to the non-normalized Thorne Stochastic Differential Equation is:

$$X = \sum_{i=1}^{n} X_0 \, w_i \, \exp\left[\mu_i - \frac{1}{2}\sigma_i^2\right](dt + \sigma_i W) \qquad (10)$$

Where:
$X$ = stochastic variable
$X_0$ = the initial value of $X$
$w_i$ = weight factor of the $i^{th}$ Gaussian
$\mu_i$ = drift coefficient of the $i^{th}$ Gaussian
$\sigma_i$ = diffusion coefficient of the $i^{th}$ Gaussian
$dt$ = time derivative
$W$ = cumulative sum of $dW$

That $X$ is the solution to the stochastic differential equation can be verified easily by calculating its first derivative, which will reproduce the original differential equation above.

**7. Determining the Thorne PDF for a given data set**

A PDF, of course, is a comprehensive description of a data set. A Thorne PDF gives an accurate description of even the most leptokurtic or skewed data set. Determining the Thorne PDF for a given data set is done semi-parametrically. It involves computing and optimizing a density histogram from the log of the raw data, or the sequential difference of such logs, then determining the values of the parameters of the Thorne PDF from that optimized density histogram.

7.1. Step one: computing an optimized histogram from a data set

Computing an optimized density histogram from a log-transformed, raw data set is a helpful precursor to determining the parameters for the log-transformed Thorne Distribution because such a histogram constrains the probability density in the tails of distributions and insures that the tails are properly represented and weighted in the final distribution. Andreotti and Douady [3] present an excellent, detailed procedure for constructing an optimized histogram. Their approach is unique in that it infers an estimate of the PDF from observed data by positing an inverse problem, then solves the inverse problem by using a regularization procedure that includes a constrained optimization. The constraints are that the smoothed PDF cannot violate canonical negentropy; that integrated probability must be conserved and must equal 1; and that likelihood must be maximized.

In the most extreme leptokurtic distributions, a stiffer curvature and weight constraint than the Andreotti procedure suggests is required in order to provide adequate smoothness in tail regions. Weights proportional to the inverse power of the local tail slope usually provide ideal smoothness without violating canonical entropy. For example, a PDF whose tails decrease locally as $x^{-(1+\alpha)}$ should be weighted as $x^{(1+\alpha)}$ in the corresponding tail regions of the smoothness functional of the objective function.

In the most leptokurtic, extreme tails, I also commonly substitute a zero-bias, kernel-density estimator. This is similar to the one elaborated by Sain [4]; but it uses, as its kernel, an estimated Thorne PDF. This Thorne kernel-density estimator uses information obtained from nearly the entire histogram to generalize suitable densities for regions where observed data is sparse. Its densities are accurate to the last, observed data point in both the positive and the negative tails. The bandwidth of the kernel is unusually broad but meets requisite zero-bias criteria. And because it is based on an estimated Thorne PDF, the kernel-density estimator permits the essential character of the Thorne Distribution to reach the tail regions of the histogram automatically, since the sparse regions of the histogram inherit the character of the kernel [5]. The Thorne PDF used for the kernel need only be a rough approximation, but it must contain sufficient leptokurtic character. An approximation is sufficient because the kernel density is dominated by bandwidth choice, which is dictated by the zero-bias criteria [6].

This new kernel-density estimator is an observation-point estimator; it, along with appropriate criteria for determining zero-bias bandwidth value in convex regions, is described in the following equation:

$$\hat{f}(x) = \frac{1}{N h_0(X_i)} \sum_{i=1}^{N} K\left(\frac{x - X_i}{h_0(X_i)}\right) = \frac{1}{N} \sum_{i=1}^{N} K_{h_0(X_i)}(x - X_i) \qquad (11)$$

Where:
$\hat{f}(x)$ = estimated probability density at the value $x$
$x$ = continuous variate value
$X_i$ = $i^{th}$ observed variate value
$i$ = counting index for the observed values of $x_i$
$N$ = number of points in the data set
$h_0(X_i)$ = bandwidth explicitly dependent upon the $i^{th}$ observed variate, as dictated by the zero-bias criteria
$K(.)$ = Thorne kernel density function that integrates to 1
$K_{h(X_i)}$ = Thorne kernel density function that incorporates bandwidth

The adaptive, zero-bias bandwidth, $h_0(X_i)$ above, is determined according to Sain's method [4].

7.2. Step two: determining the values of the parameters of the Thorne PDF from an optimized density histogram

The values of the Thorne PDF are determined by fitting Eq. 2 to the log-transformed, optimized density histogram. Specifically, this requires scaling the optimized density histogram by its smallest, positive value; adding one (1); then taking the logarithm of this sum. Scaling the histogram produces a frequency-like histogram. And adding one (1) insures that the argument of the logarithm is non-zero.

Using a least-squares methodology is a convenient means of fitting Eq. 2 to the transformed, optimized density histogram. I used the software program PeakFit 4 [7] for this purpose; but any non-linear least-squares (NLS) optimization program would suffice. I chose to work with least squares because of its long history of reliability and because of its widespread popularity in the physics and physics-related communities. Statisticians might prefer to use the popular maximum likelihood estimation (MLE) approach here and thus bypass the need for a histogram altogether. But this approach would not be appropriate in this instance because standard MLE—for normal mixture models such as a sum of Gaussian functions like the one incorporated into the Thorne Distribution—is a mathematically ill-posed problem [8]. This is due to singularities in the MLE likelihood function that occur when it is maximized and that prevent it from making accurate estimates of parameter values. These singularities result because the values of the standard deviations in the Gaussian functions of interest become infinitesimally small during optimization. In contrast, the non-linear least-squares objective function remains stable during optimization.

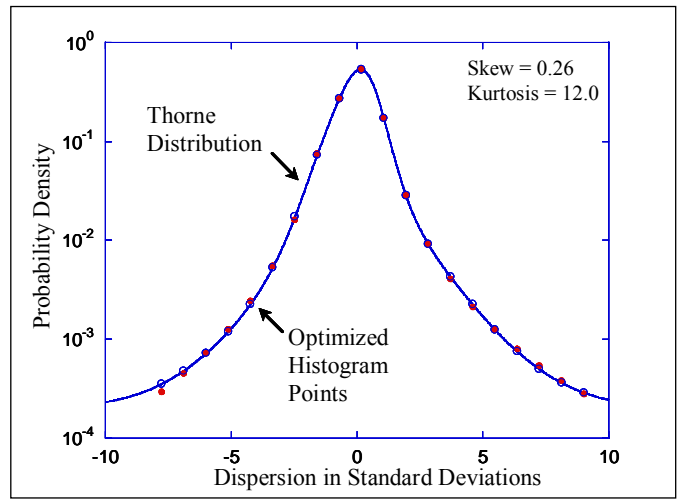

Figure 4. Values of the Thorne PDF derived from the optimized density histogram (log-return data, S&P 500 daily closing prices, 1/5/1988 to 10/10/2010) produce an accurate Thorne Distribution.

On the S&P 500 daily closing price data set shown in Figure 4, the $r^2$ was equal to 0.9994; the degrees-of-freedom adjusted $r^2$ was 0.9989; the standard error was 0.7521; and the F statistic was 2387.1. These four (4) tests were consistent with each other and, individually and collectively, they documented the accuracy of the Thorne Distribution.

The least-squares values for the parameters of the Thorne PDF, as well as their standard deviations, are guaranteed to be statistically significant when the residuals (i.e., the optimized density histogram values minus the fit values) of the fit follow a Gaussian distribution. And a delta-stabilized probability plot is useful in determining whether or not the residuals do follow a Gaussian pattern. (See Figure 5.)

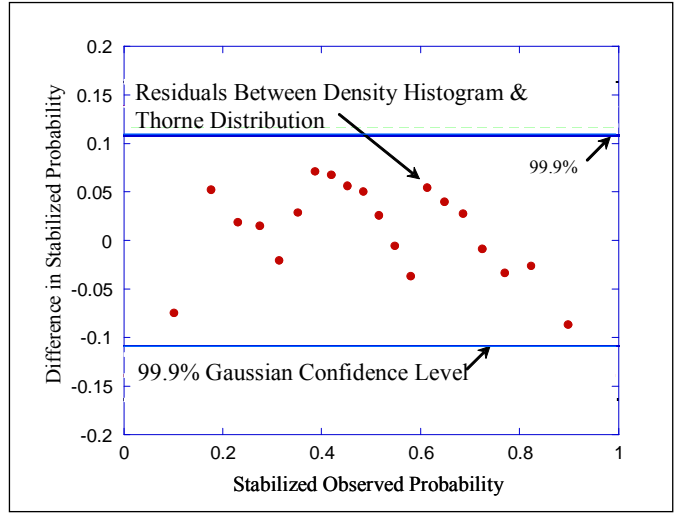

Figure 5. Thorne residuals are Gaussian per delta-stabilized [7], normal-probability plot.

7.3. A word of caution

The new distribution is a quantitative tool, but it is also almost an art form. On the first attempt, it seems cumbersome and involved. With continued use, it becomes intuitive, fast and unfailingly reliable—it behaves almost like a prescient entity. New users should be forewarned to allow for a learning curve.

**8. Simulation exercise to demonstrate the validity/correctness of the Thorne PDF through the use of a synthetic PDF**

A validation test of a new method commonly involves starting with a problem to which one already knows the answer, then determining whether or not the new method reaches the correct answer. In this instance, I chose to do a simulation exercise to validate the correctness of the new distribution. If the Thorne PDF is valid, it should be able to reproduce a synthetic PDF of a data set from variate samples drawn from that synthetic PDF. So I formulated a synthetic PDF, computed variate values from that PDF, then used those values as the starting point to calculate an optimized density histogram. Next I determined the parameter values and the optimal number of component Gaussians for the Thorne PDF by fitting the Thorne PDF, Eq. 2, to the values of the optimized density histogram.

I chose to construct a relatively difficult, synthetic PDF, one with many of the statistical characteristics of the high-frequency S&P 500 log-return PDF. Finding the underlying

PDF of the S&P 500 Index log-price return has been the passionate hope of financial research for over a decade because it has such important implications for risk measurement and control; for optimizing investment returns; for valuing financial derivatives; etc. The stylized facts which characterize it, aside from its Gaussian center, represent significant deviations from normality; they include excess kurtosis, negative skew and heavy tails with an exponent of approximately three (3) in the regions where there is power-law-like behavior.

8.1 Step One: formulating the synthetic PDF

I constructed, as a synthetic PDF, a normalized distribution, ($f_{synthetic}$), that had a Gaussian-like center and seven-halves (i.e., 7/2), power-law-like tails, as defined by Eq. 12:

$$f_{synthetic}(x) = \left[\frac{\Gamma(7/4)}{\sqrt{\pi}\ \Gamma(5/4)}\right]\frac{1}{(1+x^2)^{7/4}} \qquad (12)$$

Where:
$f_{synthetic}(x)$ = synthetic PDF used for the validation tests
$x$ = variate value
$\Gamma$ = Gamma function

The support of this synthetic distribution was (-∞, ∞); its low-order moments were: mean = 0, variance = 2, skewness = 0 and kurtosis = undefined. These moments were determined by calculating the first, four (4) central moments of the distribution by symbolic integration.

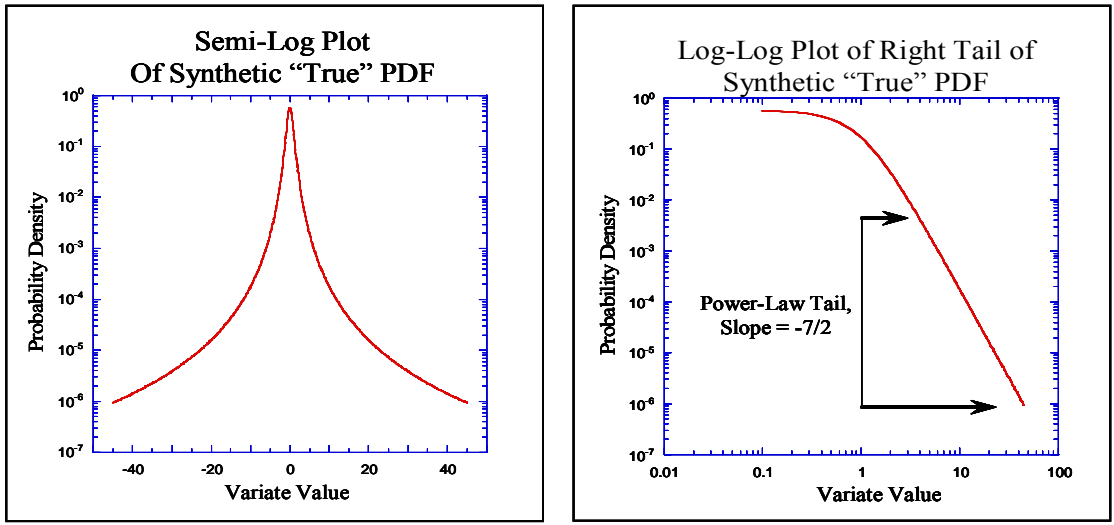

Figure 6. Two types of plots illustrate important characteristics of the formulated synthetic PDF.

The first of the plots above shows that the tails of the PDF deviate substantially from the Gaussian, which would be an inverted parabola on this graph. The second plot, which is a log-log plot, illustrates power-law tail behavior, as evidenced by the downward sloping, straight line that begins at variate values larger than about eight (8) and has a slope of close to -7/2. The synthetic PDF itself is not a power-law PDF ($f(x) = x^{-(1 + \alpha)}$), of course, since it has a quadratic, Gaussian-like center and continuous derivatives at the origin.

8.2. Step two: generating the synthetic variate values

I drew a large sample of random numbers (750,000, or the approximate equivalent of three years' S&P500 Index tick data) from the 7/2-distribution via the rejection-acceptance method [9]. This method required selecting a divisor distribution; I chose a Cauchy distribution with location = 0 and shape factor = 1/2, which gave a rejection threshold of 1.798.

8.3. Step three: computing the Thorne optimized density histogram from the synthetic variates

Computing the optimized density histogram guided the subsequent computation of the Thorne PDF. For the purposes of this simulation test, the optimized density histogram was determined from synthetic variate values; otherwise, it was computed exactly as outlined earlier in Section 7.1. The extreme tails of the optimized histogram were computed using the Thorne kernel density estimator, in conjunction with a zero-bias estimator, to set the kernel bandwidth value adaptively.

The plot below shows the optimized density histogram that was extracted from the synthetic variate values. It also shows the $f_{synthetic}$ PDF itself, for comparison. The plot shows excellent agreement between the two PDFs all the way from peak centers to far tails; thus, it clearly validates the accuracy of the Thorne optimized density histogram. Note: the plot was constructed on a semi-logarithmic graph in order to hold the tail sections of the PDFs to the most exacting standard of comparison. A linear plot would have appeared to show excellent agreement in the tail regions, even if there had been magnitudes of difference between the two (2) PDFs there, because of the extremely low tail densities.

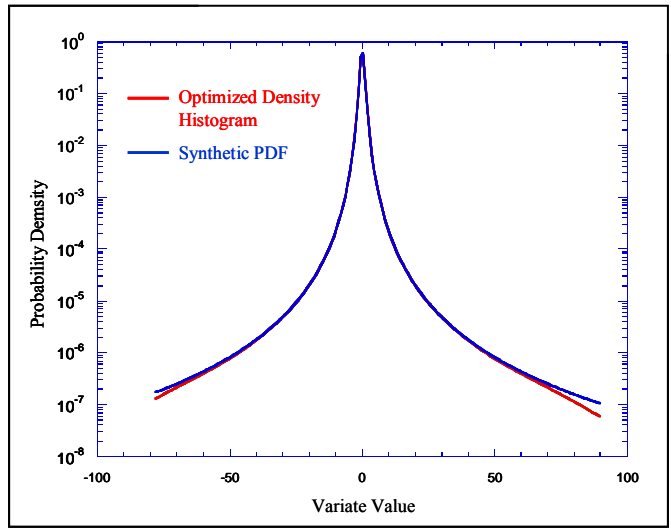

Figure 7. Thorne optimized density histogram recovers synthetic PDF, $f_{synthetic}$, reliably from synthetic variates.

8.4. Step four: fitting the equation for the Thorne PDF to the optimized density histogram

I used the software program PeakFit 4 at this stage to perform a least-squares fit of the Thorne PDF equation to the transformed Thorne optimized density histogram. (Note: the optimized density histogram is transformed by dividing all density values by the value of the minimum density and adding 1, then taking the logarithm of these values as specified in Eq. 2. This transformed density histogram is advantageous because it admits no negative values, which makes it ideal for fitting with a sum of Gaussians that can only produce positive values.) The fit then proceeded by adding component Gaussians, one at a time, until an optimal fit was achieved. Thus, successive fits provided the weights and optimized parameters of the component Gaussians, as well as their total number. The optimal number of individual Gaussians is the number that results in the largest F-statistic in the series of fits. As Figure 8 illustrates, the F-statistic of the least-squares fit increases roughly exponentially as additional, component Gaussians are added until an optimal number is reached. At that point, the F-statistic begins to decrease slowly. Once an optimal fit of the transformed density histogram has been achieved, the fit is back-transformed by reversing the transformation procedure. Since the initial, optimized histogram is properly normalized, the back-transformed fit is automatically normalized as well. Significantly, the normalization constant—as determined by numerical integration—approaches its true value asymptotically as the number of component Gaussians increases.

It takes a little practice to become proficient at specifying the initial parameters of the component Gaussians as they are added to the fit in PeakFit 4. In general, each Gaussian added requires a lower amplitude and narrower width than did those that preceded it. The entire fit is iterated successively with each new Gaussian until the number of PeakFit 4 iterations stabilizes at some small number for each added Gaussian.

Figures 8 and 9 below, along with Table 2, illustrate and detail the component Gaussians of the Thorne PDF that fit the Thorne optimized density histogram.

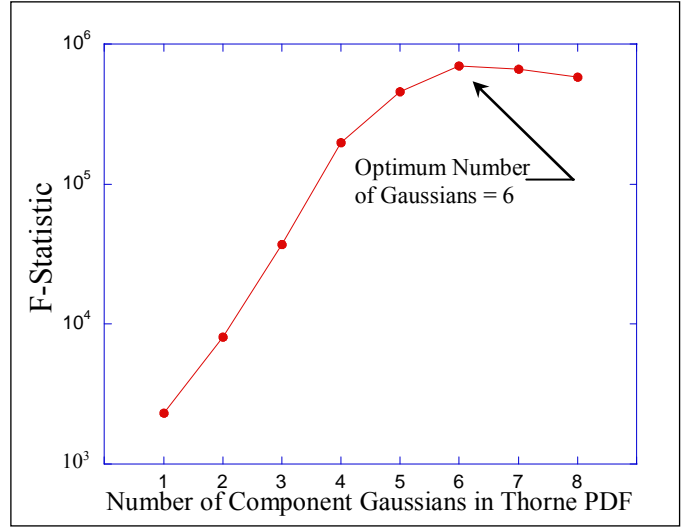

Figure 8. Representative example of optimal number of component Gaussians (i.e., 6) that corresponds to largest F-statistic (6.988 x $10^5$) in the fit of Eq. 2 to the transformed Thorne optimized density histogram.

| Peak Fit 4 Parameter Values for Component Gaussians that Fit Thorne Optimized Density Histogram | | | Peak Fit 4 *t*-Statistics for Component Gaussians that Fit Thorne Optimized Density Histogram | |
|---|---|---|---|---|
| Weight | Fixed Center | Width | Weight | Width |
| 2.41381 | 0 | 0.767862 | 10.45 | 29.16 |
| 12.2881 | 0 | 1.80448 | 26.80 | 37.08 |
| 41.7928 | 0 | 4.80233 | 31.17 | 40.03 |
| 96.2524 | 0 | 12.35919 | 24.80 | 37.91 |
| 203.2462 | 0 | 28.50726 | 28.61 | 33.93 |
| 517.6616 | 0 | 64.59965 | 57.70 | 79.22 |

Table 2. Large t-statistics verify that every parameter of the six (6), component Gaussians is essential to the close fit of the Thorne Distribution to the Thorne optimized density histogram.

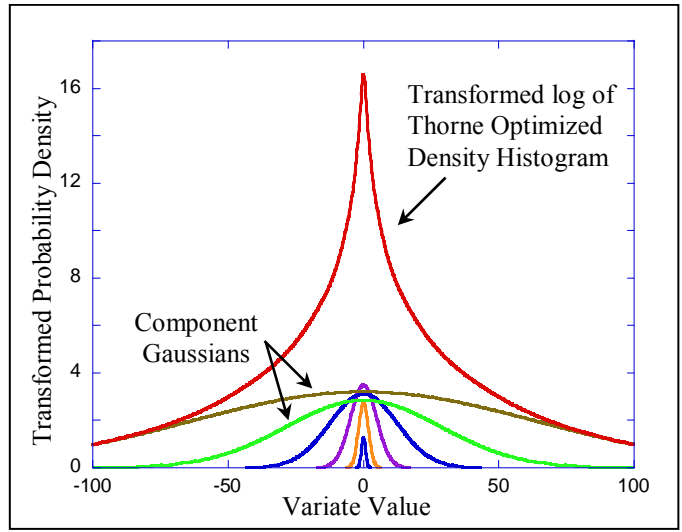

Figure 9. When summed, the six (6) component Gaussians comprise the Thorne PDF that fits the transformed Thorne optimized density histogram.

To measure the accuracy of the Thorne PDF's fit to the transformed Thorne optimized density histogram, using any of several, standard, goodness-of-fit tests is appropriate. I chose, first, to use least-squares goodness-of-fit tests, since the distribution of the residuals was Gaussian, as verified by the delta-stabilized probability plot of the residuals in PeakFit 4. The tests results were as follows: $r^2 = 0.999959$; degrees-of-freedom adjusted $r^2 = 0.999958$; standard error = 0.0330;   F statistic = 680,225 and integrated square error (ISE) = 0.1370. (This corresponds to an ISE of $3.34 \times 10^{-5}$ for an untransformed histogram.)  These values confirm a fine fit of the distribution to the optimized density histogram.  This fit was further validated as outlined in Section 8.5 below.

Incidentally, it is important to plot the final Thorne PDF and the final Thorne optimized density histogram together on both linear and semi-log plots.  The linear plot reveals any discrepancies between the PDF and the histogram in the center region, or mode; the semi-log plot reveals any discrepancies in the tail regions.  Very high kurtosis often results in a poor fit in the center of the histogram. This situation can be remedied easily by adding additional, narrow component Gaussians to the Thorne PDF.  Discrepancies in the tails can be corrected by adding broad component Gaussians.

8.5  Step five: validating the Thorne PDF for $f_{synthetic}$.

The Thorne Distribution, if it is valid, should reproduce the snythetic  PDF ($f_{synthetic}$) of the specified data set. Since both the Thorne PDF and $f_{synthetic}$ are parametric PDFs, a natural measure of accurate reproduction is the integrated square error, i.e., the integrated squared difference between the two (2) PDFs.  In the current example, the integrated square error is $5.33 \times 10^{-5}$.  This very small error provides a numerical validation of the accuracy of the Thorne Distribution.  Significantly, in this example, the difference between the Thorne Distribution and $f_{synthetic}$ is even smaller than was the difference between the distribution and the optimized density histogram, even though the least-squares fit was to the optimized histogram, not to the synthetic distribution.

Further validation of the accuracy of the Thorne Distribution can be obtained with a chi-square goodness-of-fit analysis. The chi-square test is statistically efficient, fast and broadly applicable. It is particularly appropriate for the case at hand because the residuals of the Thorne Distribution and the $f_{synthetic}$ PDF follow a Gaussian distribution, as determined by a delta-stabilized probability plot. Therefore, the sum of the squared residuals of these distributions—a value that is large for a poor fit and small for a good fit—follows a chi-square distribution.

Performing a chi-square goodness-of-fit test requires calculating the chi-square statistic of the residuals, then comparing this static to the p-value of the cumulative chi-square distribution, as specified by the appropriate degrees of freedom, i.e., the number of points at which the two distributions are compared, minus 1. The chi-square statistic in this instance was 2.26 for 20 degrees of freedom.

The chi-square analysis concluded that the null hypothesis, which is that the Thorne Distribution is statistically equivalent to $f_{synthetic}$, cannot be rejected at the $\alpha = 10\%$, 5% or 1% levels. In other words, the Thorne Distribution is a statistically valid representation of $f_{synthetic}$ of the test data set.

9. **Number of data points required to produce a Thorne PDF.**

The Thorne Distribution requires comparatively few data points; this makes it good for situations when rapid updating of a PDF is important, such as in financial analysis and density forecasting. Rapid updating can give the user early warning of impending shifts in data, like regime changes or abrupt variations in population dynamics. And such early detection of changing conditions allows the user to make timely interventions in the evolving scenario of the data set.

Its ability to utilize small data sets also makes the Thorne Distribution ideal for cluster analysis in which a population is divided into sub-groups, then modeled—via separate PDF—group by group. And the Thorne Distribution can be used as the backbone of a Thorne kernel density estimator (as described in Section 7 above) to perpetuate a consistent PDF in regions of extremely rare or even absent data points. In sum, the new distribution can access data sets that could not have been modeled feasibly before now.

To quantify the performance of the Thorne Distribution in modeling relatively small, leptokurtic data sets, I took samples (i.e., 100 points; 1,000 points; 10,000 points) of a 7/2-power-law distribution; then determined the asymptotic, mean-integrated-square-error (AMISE) [6] values for the distribution. The AMISE, of course, is a measure of error between the computed density and its corresponding, synthetic density; thus the AMISE is an appropriate measure of a model's validity. Once I had computed the AMISE values, I next determined the sample sizes required by a standard density histogram and by the respected Sheather-Jones density estimation [10] to achieve AMISE values to match those of the Thorne Distribution. The following table shows the results obtained.

| Error Measure | Number of Data Set Points Needed to Achieve Indicated AMISE | | | Ratio of Points Required for AMISE, Others vs. Thorne | |
|---|---|---|---|---|---|
| AMISE | Thorne | Sheather-Jones | Histogram | Sheather-Jones | Histogram |
| 0.0763 | 100 | 250 | 2,297 | 2.5 : 1 | 22 : 1 |
| 0.0241 | 1,000 | 3,981 | 76,634 | 4.0 : 1 | 72 : 1 |
| 0.0076 | 10,000 | 63,095 | 2,300,000 | 6.3 : 1 | 2,300 : 1 |

Table 3. New distribution leverages even small data sets into low AMISE values.

As it turned out, matching the very low AMISE values of the Thorne Distribution required many more data points (i.e., a factor of 2.5 : 1 to 6.3 : 1) with the Sheather-Jones estimator and very many more points (i.e., a factor of 22 : 1 to 2,300 : 1) with a histogram.

These results are significant because small data sets are notoriously difficult to model accurately. Small, leptokurtic or skewed data sets are even more challenging. But the Thorne Distribution can produce a valid PDF from fewer points (i.e, from much smaller data sets) than other PDF estimators can work with. Furthermore, the new distribution characterizes even very limited data sets without creating the undesirable boundary effects near the limits of the data that are often produced by other distributions.

**10. Use of the Thorne Distribution as input to coherent-risk measures**

Most current measures of coherent risk [11] require characterization of tail probability densities. Either the Generalized Extreme Value distribution or the Pareto distribution is commonly used to provide such tail-density input. Extreme Value Theory (EVT) is used to determine the parameter values for these distributions. But the Generalized Extreme Value (exponential algebraic tail) and Pareto (algebraic tail) distributions are suitable for nothing but tails, and not even all of those; other distributions must be patched in to model the remaining sections of a PDF. And the myriad, documented [1] weaknesses of EVT contaminate the densities produced by the input distributions (GEV and Pareto).

The Thorne Distribution is a more direct and reliable source of input to such measures of coherent risk as Expected Shortfall and Maximum Drawdown. The parameters of the Thorne Distribution are based on a highly-accurate optimized histogram; the distribution models the entirety of the most challenging PDFs; and the distribution can adapt from algebraic to exponential tails as required by data. For all of these reasons, the new distribution has an intimate connection with observed data that can contribute to precise, reliable, robust assessments of coherent risk.

**11. Additional tools built on the new distribution**

To test potential breadth of application, I expanded on the basic Thorne Distribution and its underlying stochastic differential equation by continuing development that ultimately included the kernel-density estimator discussed herein, as well as a joint distribution, a copula, a stochastic-shock string model, a coherent-risk measure and a large suite of exclusively financial-market-related tools, including a stunning trade-timing indicator. The fundamental Thorne Distribution, with its attendant accuracy and strength, enables each of these applications and is, in turn, propagated through them.

**12. A wider perspective on the practical applicability of the Thorne Distribution**

This paper has referred repeatedly to the challenge posed by fat-tailed data in the financial world and to the apparent power of the Thorne Distribution to meet that challenge. The distribution's accurate characterization and modeling of notoriously leptokurtic S&P 500 data was illustrated, for instance, which indicated its value as highly accurate input to, among other things, greatly improved coherent-risk measures. But the distribution may also hold important promise in stock selection; in portfolio formation and optimization; in the valuation of derivative instruments such as options and futures; in the estimation of liquidity requirements for institutions; in the quantification of price volatility; in probability price-range forecasting; in the calculation of marginal distributions in copula-based, multivariate risk models; in the generation of stochastic paths for Monte Carlo simulations of price paths in continuous-time models, including those for look-back and exotic options and structured instruments; etc.

And fat-tail issues are not restricted to Wall Street; they manifest in the statistical behaviors of countless phenomena, systems and environments—in all of which the Thorne Distribution should be applicable. Such situations might range widely from internet traffic statistics (including those of queuing systems, TCP files, packet-transfer rates/sizes and session lifetimes), for instance, to the turbulence characteristics in aerodynamics/hydrodynamics, to the epidemiology of disease outbreaks.

**13. A wider perspective on the theoretical implications of the distribution**

The Thorne Distribution could be written into production code to begin to play an immediate, positive role in improving the handling of leptokurtic data dramatically in a variety of workplace settings. But it may have theoretical implications that warrant further thought too.

A comprehensive discussion of the Thorne Distribution should perhaps include a consideration of its fractal properties and of its probable relevance to the work of Benoit Mandelbrot and the topics of self-similarity and/or emergent and self-organizing systems. The distribution is fundamentally self-similar, since its logarithm is composed of Gaussians that differ only in scale (standard deviation) and translation (mean); and taken together, these component Gaussians describe multi-scale probability density.

Even more basically, the Thorne Distribution has about it a sense of something innate, something fundamental. It encompasses, as a sub-set, the time-honored Gaussian distribution. And it is itself a comprehensive distribution of distributions capable of characterizing empirically--not just the critical realities of tail behavior on many scales–but literally the full extent of the practical probability space of any data set. This is an enormous analytical advantage. And in addition, the individual component Gaussians of the Thorne Distribution may, in some circumstances, characterize important sub-processes or sub-populations within a larger process of which a given data set is only a single realization.

## 14. Conclusion

I have proposed a new distribution to quantify and resolve the fat tails associated with leptokurtic and skewed data; outlined the mathematical and statistical constructs of this distribution; and demonstrated its accuracy and validity. The Thorne Distribution may be a door into a new era in statistical analysis/modeling in finance and elsewhere.

## Acknowledgments

My thanks to N. M. Thorne for significant contributions to this paper.